\documentstyle{article}
\begin{document}

\title{Everyone Makes Mistakes $-$ Including Feynman
}

\author{Toichiro Kinoshita\\
Newman Laboratory, Cornell University,\\
Ithaca, New York 14853}

\maketitle

This talk is dedicated to Alberto Sirlin in celebration of
his seventieth birthday.
I wish to convey my deep appreciation of his many important
contributions to particle physics
over 40 years and look forward to many
more years of productive research.

\section{Introduction}

Alberto arrived at Cornell as a graduate student
 in September, 1955.
I had come to Cornell as a research associate a few months earlier.
Thus I have been acquainted with him for 45 years.

It was the time when experimental observations
of some weak interaction processes, 
the so-called $\theta - \tau$ puzzle in particular,
began to expose the internal inconsistency
in the theory of the weak interaction.
Analyzing this problem in great depth,
Lee and Yang concluded that parity conservation,
assumed to be valid in previous theories of
the weak interaction, was the most likely culprit. 
They suggested that parity symmetry is not valid 
in the weak interaction and proposed ways to test it 
experimentally \cite{ly1}.
The experimental verification followed soon afterward \cite{wu,garwin}.
The two-component neutrino theory (discarded previously by Pauli) 
became the favored theory \cite{ly2,salam,landau}.

\section{Radiative correction to muon decay}

A detailed comparison of the 2-component theory with experiment would
require accounting for radiative corrections.
This is because they might be as large as
$\alpha \omega^2 \simeq 28.4/137$, where   
$\omega = \ln (m_\mu / m_e ) = 5.3316$, according to the paper
\cite{behrends}
on radiative corrections
to the parity-conserving muon decay, 
on which Alberto worked
before he came to Cornell.
Its extention to the parity-non-conserving case 
is not difficult.  Thus Alberto's
experience enabled us to jump-start the calculation
of the parity-non-conserving muon decay 
and finish it on very short notice.
The radiatively corrected muon decay spectrum we obtained
in the two-component neutrino theory is 
\cite{ks1}
\begin{eqnarray}
dN_r (x, \theta )&=&{1 \over 2} A [ 3-2 x + {\alpha \over {2\pi}} f(x)
+ 6 \zeta {m_e \over m_\mu } {{1-x} \over x}   \nonumber  \\ 
 &+&\xi \cos \theta \{ 1-2 x + {\alpha \over {2\pi}} h(x) \}
 ] x^2 dx d \Omega ,
\end{eqnarray}
where $A, \xi, \zeta$ are functions of 
weak coupling constants, $x= 2p_e /m_\mu$, and
\begin{eqnarray}
f(x)&=&(6-4x) u(x) + (6-6x) \ln x   \nonumber  \\
&+&{{1-x} \over {3x^2}} [ (5+17x-34x^2)(\omega + \ln x) 
-22x+34x^2 ],  \nonumber  \\
h(x)&=&(2-4x) u(x) + (2-6x) \ln x  \nonumber  \\
&+&{{1-x} \over {3x^2}} [ (-1-x-34x^2)(\omega + \ln x)  \nonumber  \\
&-&3+7x+32x^2 - {{4(1-x)^2} \over x} \ln (1-x) ] ,
\label{fandh}
\end{eqnarray}
with
\begin{eqnarray}
u(x)&=&\omega^2 + \omega ({1 \over 2} -2 \ln 2) + 2 \ln 2 -3 
+ (2\omega -1 - {1 \over x} ) \ln (1-x)  \nonumber  \\
&+& \ln x \left [ 3 \ln (1-x) - \ln x - 2 \ln 2 \right ] 
+ L(1) - 2 L(x),
\end{eqnarray}
and 
\begin{equation}
L(x) = \int_0^x \ln (1-t) (dt/t) ,~~~~ L(1) = - {\pi^2 \over 6} .
\end{equation}
The radiative correction to the decay lifetime is large,
being proportional to $\omega^2$:
\begin{equation}
~~~~~{{\tau - \tau_0} \over \tau_0 } = - {\alpha \over {2\pi }} ( \omega^2 + \cdots) \simeq - 0.02973 .
\end{equation}

We presented our result at the Rochester conference 
in the spring of 1957.
Then, one day in 1958, lightening struck us.
We received a preprint from Berman
stating that he  disagreed with our result.
In particular, he mentioned that the radiative correction to 
the muon lifetime is linear, not quadratic, in $\omega$,
contrary to our result.
When we read his preprint, however, we suspected immediately that
his result, which contains a term linear in $\omega$,
 must also be wrong.
Alberto and I worked hard for a week or two and found
that our intuition was in fact correct. 

In the new result \cite{ks2},
spectral functions $f(x)$ and $h(x)$ have the same form as 
in (\ref{fandh}), but
 $u(x)$ is replaced by
\begin{eqnarray}
R(x)&=& \omega  [ 1.5 + 2 \ln (1-x) - 2 \ln x ] 
- 2 L(x) + 2  L(1) - 2  \nonumber \\
&-& \ln x (2\ln x -1) + ( 3 \ln x -1 -{1 \over x} ) \ln (1-x).
\end{eqnarray}
The decay spectrum still contains terms linear in 
$\omega (\equiv \ln (m_\mu / m_e ))$. 
However, the muon decay lifetime now has no dependence at all on $\omega$
and the net radiative correction is very small: 
\begin{equation}
~~~~~{{\tau - \tau_0} \over \tau_0 } = - {\alpha \over {2\pi }} 
\left ( {25 \over 4} - \pi^2 \right )  \simeq  4.17 \times 10^{-3} .
\end{equation}

To discuss our mistake in \cite{ks1} that Berman pointed out,
let me focus on the inner bremsstrahlung contribution to
the muon decay, which contains the factor
\begin{equation}
\sum_i \int_0^{k_{max}} {{d^3 k} \over \epsilon}
\left [ {{p_2 .e_i} \over {p_2 .\kappa}} 
- {{p_1 .e_i} \over {p_1 .\kappa}} \right ]^2  ,
\label{polsum}
\end{equation}
where $p_1 , p_2$, and $\kappa$ are the 4-momenta of 
the muon, electron, and photon,
and $e_i$ are the polarization vectors of the photon.

Recall that, in a covariant calculation, the virtual photon 
is often treated as a vector meson of mass $\lambda$
with the understanding that $\lambda \rightarrow 0$
in the physical limit.
To be consistent with this, the
{\it real photon} must also  be regarded as a vector meson of mass $\lambda$.
This means, in particular, that the sum in (\ref{polsum}) must be carried out 
over four polarizations including time-like and longitudinal polarizations.
Our mistake was that we had summed only over transverse polarizations.
Furthermore, even
in the limit $\lambda \rightarrow 0$,
the contribution of non-transverse polarizations does not vanish
if the photon has infrared divergence. 
As a matter of fact, this extra contribution has an $\omega$
dependence that cancels the leading $\omega$ dependence 
from transverse photons.

This was supposed to be well-known: it 
is related to Feynman's famous error
in matching non-relativistic and relativistic calculations 
of the Lamb shift that was discovered by French and mentioned
in Footnote 13 of Feynman's paper \cite{feynman}.
Unfortunately, many people, including us, 
had forgotten or failed to appreciate the significance of
his footnote and made the same mistake again and again.
(It is true that the connection with our mistake is somewhat obscure since
Feynman's footnote does not deal directly with scattering states
or decaying states.)
In the end Berman agreed with us
and revised his paper  accordingly \cite{berman1}.

The lessons we learned from this episode are:

$\bullet$ 
Although the differential spectrum of $\mu - e$ decay
diverges logarithmically as $m_e /m_\mu \rightarrow 0$,
the {\it total} decay rate is finite in this limit.

$\bullet$ 
Cancellation of infrared divergences
 is a necessary but not sufficient
guarantee for the computation to be correct.
(Many people were unaware of this 
and made the same mistake, even after our paper was
published.)

We obtained a similar result for nuclear $\beta$ decay
under some simplifying assumptions.
The radiative correction to the
$\beta$-ray spectrum in the V - A theory 
(for $m_e \ll E$) is found to be \cite{ks2}
\begin{eqnarray}
\Delta P d^3 p &=& {\alpha \over \pi^4} G^2 E_m^5 (1-x)^2 x^2 dx   \nonumber  \\
&~&\left \{ 6 \ln \left ( {\Lambda \over m_p} \right )
       + 3 \ln \left ( {m_p \over {2E_m}} \right )
       + {3 \over 2} - {{2\pi^2 } \over 3}
\right .   \nonumber  \\
&~&\left .
+4 (\ln x -1) \left [ {{1-x} \over {3x}} - {3 \over 2}
+ \ln \left ( {{1-x} \over x} \right ) \right ]
+ {{(1-x)^2} \over {6x^2}} \ln x \right .  \nonumber  \\
&~&\left .
+ \Omega \left [ {{4(1-x)} \over {3x}} -3 + {(1-x)^2 \over {6x^2}} 
+ 4 \ln \left ( {{1-x} \over x} \right ) \right ] \right \} , 
\end{eqnarray}
where $ x=E/E_m$, $\Omega = \ln (2E_m / m_e )$,
$E_m$ is the maximum total energy of the electron,
and $m_p$ is the proton mass.
The spectrum diverges logarithmically for $m_e \rightarrow 0$.
However, the correction to the lifetime has no $\ln m_e$ dependence:
\begin{equation}
{{\Delta \tau} \over \tau_0} =
-{\alpha \over {2\pi}} \left [ 6 \ln \left ( {\Lambda \over m_p} \right )
+ 3 \ln \left ( {m_p \over {2E_m}} \right )  - 2.85 \right ] .
\end{equation}
Thus $\beta$ and $\mu$ decays
share the same $\ln m_e$ dependence, 
suggesting that it is 
more general and not limited to these decays.
(See Appendix A.1, too.)

Feynman came to Cornell in the fall of 1958 for three months.
He explained to me how he and Berman made exactly the same mistake.
Feynman had asked Berman to check our calculation
for his thesis work.
But, actually, Feynman himself was doing this calculation
independently of Berman.
At the end they compared notes
and were satisfied that their results agreed.
When confronted with our new result which differed from theirs,
they checked the notes once again and found that
they made the same mistake in copying the bottom equation of a page
to the top of next page.
Feynman was so disturbed by this mistake
that he told me how sorry he was more than a few times while
at Cornell.

\section{Radiative correction to $\pi -e$ decay}

Feynman brought with him
a new preprint of Berman on the radiative correction to $\pi - e$ decay.
As is well-known, 
in the $V-A$ theory,
the ratio of $\pi -e$ and $\pi -\mu$ decay rates is
\begin{equation}
R_0 = \left ( {m_e \over m_\mu } \right )^2
 \left ( {{m_\pi^2 - m_e^2} \over {m_\pi^2 - m_\mu^2 }} \right )^2
 \simeq 1.28 \times 10^{-4}
\end{equation}
if the radiative correction is not included.
Berman's result, including one loop radiative correction, was of the form 
\begin{equation}
R = R_0 \left ( 1 - {{3\alpha} \over \pi} \ln (m_\mu / m_e ) + \cdots \right ) .
\end{equation}
This $R$ has a rather large correction ( $\sim 3 \%$ )
and looked strange since $R/R_0$ diverged for $m_e /m_\mu \rightarrow 0$.
Feynman and I were so puzzled by this result, 
which seemed to contradict what was discovered 
in $\mu$ and $\beta$ decays,
that we decided to check it with a fresh calculation.
For the next two months we worked hard, totally
independently of each other, except that we agreed to start from the
same effective Lagrangian 
\begin{equation}
~~~~~~~~~~~~~g m_{\it l} \bar{\psi}_{\it l} a \psi_\nu \phi_\pi  ,
\label{wronglagrangian}
\end{equation}
where $l$ represents either a muon or an electron and 
$ a = (1 + i\gamma_5 )/2$.

The radiative correction due to the virtual photon is straightforward
\cite{kinoshita1}:
\begin{eqnarray}
{{\Delta P_{VP}} \over P_0} &=&
 {\alpha \over \pi} \left [ {3 \over 2} \ln {\Lambda \over m_\pi}
- b(r) \left ( \ln { \lambda_{min} \over m_\pi } - {1 \over 2} \ln r + {3 \over 4} \right )    
+ {r^2 \over { 1 - r^2}} \ln r + {1 \over 2} \right ] ,
\nonumber  \\
&~&~~~~~~~~~~~~~b(r) = 2 \left ( {{1+r^2} \over {1-r^2}} \ln r + 1 \right ) ,
\label{VP}
\end{eqnarray}
where $P_0$ is the uncorrected $\pi - e$ decay rate,
$r=m_e /m_\pi$, 
$\lambda_{min}$ is the infrared cutoff mass,
and $\Lambda$ is the ultraviolet cutoff mass.

The total probability of the inner bremsstrahlung correction 
is \cite{kinoshita1}
\begin{eqnarray}
{{\Delta P_{IB}} \over  P_0} &=&
 {\alpha \over \pi} \left [ 
b(r) \left ( \ln { \lambda_{min} \over m_\pi } - 
\ln (1-r^2 ) - {1 \over 2} \ln r + {3 \over 4} \right )
\right .   \nonumber  \\
& & \left . - {{r^2 (10 -7 r^2 )} \over {2 ( 1-r^2 )^2}} \ln r
+ {{2 (1+r^2 )} \over {1-r^2 }} L (1-r^2 )
+ {{15-21r^2 )} \over {8( 1-r^2 )}} \right ]  .
\label{IB}
\end{eqnarray}
It follows from (\ref{VP}) and (\ref{IB}) that,
to order $\alpha$, the rate of $\pi -e$ decay is
\begin{equation}
P = P_0 (1 + \eta_e )  ,
\end{equation}
where 
\begin{eqnarray}
 \eta_e &=& {\alpha \over \pi} \left [ 
{3 \over 2} \ln \left ( \Lambda \over m_\pi \right )
- b(r) \ln (1-r^2 ) 
- {{r^2 (8 -5 r^2 )} \over {2 ( 1-r^2 )^2}} \ln r
\right .   \nonumber  \\
& & \left . 
+ {{2 (1+r^2 )} \over {1-r^2 }} L (1-r^2 )
+ {{19-25r^2 } \over {8( 1-r^2 )}} \right ] .
\label{eta_e}
\end{eqnarray}
Infrared divergences have canceled out as expected.

When we finished the calculation, we compared the results
and found that we agreed with each other.
Unfortunately, we did not agree with Berman.
In particular, our result did not have the $\ln (m_\mu / m_e )$ term.
We thought for a while that Berman's calculation
was wrong.
But, after scrutinizing our calculation very closely, 
I realized that it was we that were wrong.
We committed a very subtle mistake:
it was in our choice of the starting Lagrangian.

Berman started from an effective Lagrangian with the derivative coupling:
\begin{equation}
~~~~~~~~~~~~~g  \bar{\psi}_{\it l} \gamma_\mu  a \psi_\nu 
( i{{\partial \phi_\pi} \over {\partial x_\mu }}
-e A_\mu \phi_\pi )  .
\label{bermanlag}
\end{equation}
To simplify the algebra this is often turned into
a non-derivative form 
\begin{equation}
~~~~~~~~~~~~~g (m_{\it l} - m_\nu ) \bar{\psi}_{\it l} a \psi_\nu \phi_\pi
\label{equivlag}
\end{equation}
by integration by parts and
use of the equation of motion
\begin{equation}
i {\partial \over {\partial x_\mu }} \bar{\psi_{\it l}} \gamma_\mu 
+ eA_\mu \bar{\psi_{\it l}} 
+ m_{\it l} \bar{\psi_{\it l}}  = 0 .
\label{eqmo}
\end{equation}

Our Lagrangian
(\ref{wronglagrangian})
was obtained from the equivalent Lagrangian (\ref{equivlag})
assuming that $m_\nu = 0$ and $m_{\it l}$ is the physical mass.
Unfortunately, we did not realize initially that
this equation is not valid to order $e^2$
if $m_{\it l}$ is the physical mass.
The correct equation requires the self-mass term 
$\delta m_{\it l} \bar{\psi_{\it l}} $ on the right-hand side 
of (\ref{eqmo}).
Or, equivalently, we may rewrite the corrected equation as
\begin{equation}
i {\partial \over {\partial x_\mu }} \bar{\psi_{\it l}} \gamma_\mu 
+ eA_\mu \bar{\psi_{\it l}} 
+ m_{\it l}^0 \bar{\psi_{\it l}}  = 0 ,
\end{equation}
where $m_{\it l}^0$ is the {\it bare} mass.
Berman's Lagrangian (\ref{bermanlag}) may thus be replaced by 
\begin{equation}
~~~~~~~~~~~~~g m_{\it l}^0 \bar{\psi}_{\it l} a \psi_\nu \phi_\pi  .
\end{equation}
This means that we can turn our incorrect result into a correct one
by simply replacing the renormalized mass with the bare mass.
The appearance of bare mass in this context had been noticed
by Ruderman \cite{ruderman,gatto}.

The radiatively corrected decay ratio $R$ 
can thus be written as \cite{kinoshita1}
\begin{equation}
~~~~~~~R = R_0 ((1 + \eta_e )/(1+ \eta_\mu ))(1 + \delta )
\end{equation}
where $\eta_e$ and $\eta_\mu$ are defined by (\ref{eta_e}) and
\begin{eqnarray}
~~~~~ \delta &=&  ( m_e^0 / m_e )^2/(m_\mu^0 /m_\mu )^2 -1   \nonumber  \\
~~~~~~~~~~~&=&  - (3\alpha /\pi ) \ln (m_\mu / m_e )    \nonumber  \\
~~~~~~~~~~&~&\simeq  - 15.995~ (\alpha /\pi ) .
\end{eqnarray}
This  $R$ is in exact agreement with Berman's result \cite{berman2}.

Measurement of the $(\pi -e) /(\pi -\mu )$ decay ratio was just starting
at the time of this calculation.
The experimental uncertainty was still so large that
the presence or absence of the $(1+\delta)$ factor could not be tested
experimentally.
Later more accurate measurements confirmed this large effect \cite{britton,czapek}
\begin{eqnarray}
~~~~~~~R &=& 1.2265~(34)~(44) \times 10^{-4} ,   \nonumber  \\
~~~~~~~R &=& 1.2346~(35)~(36) \times 10^{-4} .
\end{eqnarray}

Our calculation of $R$ relied on
an implicit untested assumption that the UV cutoff $\Lambda$
is common to both $\pi - e$ and $\pi - \mu$ decays.
A justification of this assumption had to wait for 
the emergence of the Standard Model \cite{ms1}.
The hadronic effect was also taken into account \cite{ms2}.
This leads to the latest value 
\begin{equation}
~~~~~~~R = 1.2352~(5) \times 10^{-4} .
\end{equation}

Working with Feynman was a very interesting and instructive experience.
Let me share one episode with you.

As is well-known, the integration over 3-body final states
is quite non-trivial.  
I spent most 
of the two months checking my calculation of the inner bremsstrahlung
term $\Delta P_{IB}$ again and again.
In the end
more than 30 pages of equations
were needed to carry my conventional approach to the end.
It also took about two months for Feynman to
evaluate this integral.

Actually, I am not sure that he was working on this problem
all the time.
His office was next to mine
so that I could hear that  
he was constantly practicing bongo drums 
using the cover of the heating system
as a drum.
When we finally finished the work and compared notes, however,
I was astounded to find that his whole calculation
was written on just two sheets of paper.
What he was doing during the two months was 
not only playing bongo but also looking for new ways of doing
the integration.
And he actually found a very simple and elegant method !

Since this does not seem to be widely known,
let me describe it here.
The  decay process $\pi \rightarrow e + \bar{\nu} + \gamma$
 has 4-momentum conservation:
\begin{equation}
p_\pi = p_e + p_\nu + k  .
\end{equation}

\noindent
$\bullet$ Step 1:
Take any final-state $p_e$ and go to the reference frame in which
the space-components of $p_\pi$ and $p_e$ satisfy the relation
\begin{equation}
\vec{p}_\pi = \vec{p}_e .
\end{equation}
Then $\vec{p}_\nu$ and $\vec{k}$ are exactly back to back.
Thus the angular integration becomes trivial.
The result is a function of the fourth component $E_e$ of $p_e$ only,
which can be easily converted to a covariant form.

\noindent
$\bullet$ Step 2:
Go to the pion rest frame by an $E_e$-$dependent$
Lorentz transformation.
Then the remaining integration over $E_e$ is almost trivial.
That's all.

Before we finished our work, Feynman went back to Caltech. 
After some exchange of letters
discussing fine details of the calculation 
and the drafting of a report, Feynman told me
to publish the paper by myself, which I did reluctantly \cite{kinoshita1}. 
He did not explain why he did not want to put his name on it.
I can only guess some reasons.
One is that he was not comfortable with the appearance
of unphysical mass in observable quantities.
Since the Lagrangians used in these days were just effective Lagrangians
and not renormalized ones,
they did not provide a proper framework to deal with such a problem.
Only within the context of renormalizable theories, such as the Standard Model,
can one treat it properly in terms of the renormalization group.

Although the presence of the $\ln (m_\pi /m_e )$ term
in the $\pi - e$ decay rate 
seemed strange at first sight, it was actually
not so strange.
This is because the $\pi -e$ decay amplitude in the $V-A$ theory
is proportional to $m_e$,
which multiplies $\ln (m_\mu /m_e )$
and makes the whole amplitude vanish for $m_e \rightarrow 0$.
Note also that the $\eta_e$ part of the radiative correction
to the $\pi - e$ decay
behaves in the same manner as those of $\mu$ and $\beta$ decays,
namely, it has no $\ln m_e$ term.

\section{Mass singularity}

These examples convinced me that the striking cancellation of
$\ln m_e$ terms in the total probability
is a very general feature of quantum field theory.
It seemed that the structure of general Feynman amplitudes
in the massless limit deserved some attention.
I spent the next few years trying to understand this problem.

The basic tool of analysis was the power counting rule to examine 
the behavior of Feynman integrals in the zero mass limit.
As a function of (various) masses, the Feynman amplitude has
a very complicated structure at zero-mass points.
In general its value depends on the order and direction in which
the zero-mass points are approached.
It is found, nevertheless, that
$m \rightarrow 0$ in a propagator of mass $m$
does not cause divergence unless it is enhanced by
putting vertex-sharing propagators (including external lines)
on the mass shell.
The result was reported in \cite{kinoshita2}.
Alberto used it in his derivation of 
differential equations  for propagators and
vertex functions in QED, and obtained results
which are equivalent to the Callan-Symanzik equation \cite{sirlin}.
(See Appendix A. 3.)

\section{Lepton anomalous magnetic moments}

It turned out that the analysis of the mass singularity was very useful
in studying the $m_\mu /m_e $ dependence of the
muon anomalous magnetic moment \cite{kinoshita3}.
This was in fact the beginning of my active involvement in the
$g-2$ problem, from which I have not yet managed to extract myself.
The paper \cite{kinoshita3} was extended to the study of 
higher-order muon anomalous moment
based on the renormalization group technique \cite{lautrup,kno1}.

The analytic tool developed in \cite{kinoshita2}
to deal with general Feynman-parametric integrals
also turned out to be very handy as
the starting point of my
work on the sixth- and eighth- order
radiative corrections to the
lepton $g-2$ by a numerical method \cite{ck1,lk1}.
After converting momentum space Feynman integrals
into Feynman-parametric integrals analytically,
we evaluated them numerically 
using the iterative-adaptive Monte Carlo
integration routine VEGAS \cite{lepage}.

In the case of the electron $g-2$,
the best value of the coefficient of $(\alpha/\pi )^3$, 
obtained by VEGAS, is \cite{kinoshita4}
\begin{equation}
~~~~~~~A_6^{(num)} = 1.181~259~(40) .
\end{equation}
This is in good agreement with the analytic result
obtained by Laporta and Remiddi several months later,  
after many years of hard work 
\cite{laporta}:
\begin{equation}
~~~~~~~A_6^{(anal)} = 1.181~241~456 \cdots .
\end{equation}
  
At present $A_8$,
the coefficient of $(\alpha/\pi )^4$, 
is known only by the VEGAS integration.
The most recent reported value of $A_8$ 
is \cite{hughes}
\begin{equation}
~~~~~~~A_8^{(num)} = - 1.509~8~(384) .
\end{equation}
The project to reduce the
uncertainty of $A_8$ by a factor of 3 or more 
by means of massively-parallel computers is  
approaching the final stage.

At present the best theoretical value of $a_e$,
including small electroweak and hadronic terms, is 
\begin{equation}
a_e (th) = 1~159~652~153.5~(1.2)~(28.0) \times~ 10^{-12}
\label{aeth}
\end{equation}
evaluated using the $\alpha$
obtained from the quantum Hall effect \cite{jeffery,mohr}:
\begin{equation}
\alpha^{-1} (qH) = 137.036~003~7~(33) .
\label{quantumhall}
\end{equation}
The value $\pm 1.2$ in (\ref{aeth}) is the remaining uncertainty in theory.
The result (\ref{aeth}) is to be compared with the
measured values of $a_e$ obtained in Penning trap experiments \cite{vandyck1}:  
\begin{eqnarray}
a_{e^-} &=& 1~159~652~188.4~(4.3) \times 10^{-12},  \nonumber  \\
a_{e^+} &=& 1~159~652~187.9~(4.3) \times 10^{-12},
\end{eqnarray}
or their weighted average \cite{mohr}
\begin{equation}
~~~~~~~~~a_e = 1~159~652~188.3~(4.2) \times~ 10^{-12}.
\end{equation}
Theory is - 1.3 standard deviations away from experiment.

The QED contribution to $a_\mu$ has been computed
through five loops\cite{km,cm}
\begin{eqnarray}
a_\mu ({\rm QED}) &=& 0.5 \left ({\alpha \over \pi} \right )
+ 0.765~857~376~(27) \left ( {\alpha \over \pi} \right )^2 
+ 24.050~508~98~(44) \left ( {\alpha \over \pi} \right )^3
\nonumber   \\
&+& 126.07~(41) \left ( {\alpha \over \pi} \right )^4
+ 930~(170) \left ( {\alpha \over \pi} \right )^5  \nonumber   \\
&=& 116~584~705.7~(2.9) \times~ 10^{-11}.
\label{amuqed}
\end{eqnarray}
The coefficients of $(\alpha/\pi)^n$ are mass-dependent.
Many $\ln (m_\mu /m_e )$  terms as well as some mass-independent
terms can be determined analytically by renormalization group
considerations \cite{kinoshita3,lautrup,kno1}.
Coefficients of $\alpha^2$ and $\alpha^3$ 
can be evaluated to any precision by expansion in mass ratios \cite{cm}.
The errors in the $\alpha^2 $ and $\alpha^3$ terms 
come only from measurement uncertainties of $m_e /m_\mu$
and/or $m_e /m_\tau$.
The coefficient of $\alpha^4$ is known by numerical integration
only. The coefficient of $\alpha^5$ is only a rough estimate
at present \cite{km,karsh}.
The electroweak contribution has been evaluated to two-loop order
\cite{cm}
\begin{equation}
a_\mu ({\rm EW}) = 152~(4) \times~ 10^{-11}.
\label{amuew}
\end{equation}
The  current best estimate of the hadronic contribution is \cite{cm}
\begin{equation}
a_\mu ({\rm had}) = 6739~(67) \times~ 10^{-11}.
\label{amuhad}
\end{equation}
The sum of (\ref{amuqed}), (\ref{amuew}), and (\ref{amuhad}) gives the 
prediction of the Standard Model
\begin{equation}
a_\mu ({\rm theory}) = 116~591~597~(67) \times~ 10^{-11}.
\end{equation}
This is in good agreement with the value 
\begin{equation}
a_\mu ({\rm exp}) = 116~592~050~(460) \times~ 10^{-11} 
\end{equation}
obtained by combining the CERN result and the data taken
through 1998 at Brookhaven National Laboratory \cite{cm,brown}.

\section{Fine structure constant  as test of quantum mechanics}

As is seen from (\ref{aeth}), the uncertainty in $a_e (th)$
is dominated by that of $\alpha$ given in (\ref{quantumhall}).
This means that this $\alpha$ is not accurate enough to test QED 
to the extent allowed by
the precision of the measurement and theory of $a_e$.
The situation is no better for other 
high precision values of $\alpha$ determined from 
the ac Josephson effect \cite{mohr},
measurement of $h/m_n$ ($m_n$ is the neutron mass) \cite{krueger},
muonium hyperfine structure \cite{liu},
and Cesium $D_1$ line \cite{udem,haensch}:

\begin{equation}
\alpha^{-1} (acJ\& \gamma_p^{'} ) = 137.035~988~0~(51) ~~~[3.7 \times 10^{-8}] ,
\end{equation}

\begin{equation}
\alpha^{-1} (m_n) = 137.036~011~9~(51) ~~~~~~[3.7 \times 10^{-8}] ,
\end{equation}

\begin{equation}
\alpha^{-1} (\mu hfs) = 137.035~993~2~(83)~~~~[6.0 \times 10^{-8}] ,
\end{equation}

\begin{equation}
\alpha^{-1} (C_s D_1) = 137.035~992~4~(41)~~~~[3.0 \times 10^{-8}] .
\end{equation}
Atom beam interferometry, single electron tunneling,
fine structure of the helium atom,
and bound electron $g-2$,
may also produce very precise values of $\alpha$.

This means, however, that it is
the electron $g-2$ that can provide the most precise 
value of $\alpha$ at present. 
>From the Seattle experiment and QED one obtains
\begin{eqnarray}
\alpha^{-1} (a_e) &=& 137.035~999~58~(14)~(50)~~   \nonumber  \\
&=& 137.035~999~58~(52)~~~~ [3.8 \times 10^{-9}].
\end{eqnarray}
Errors on the first line are due to the $\alpha^4$ term and 
measurement of $a_e$.

When new experiments 
are completed, the measurement uncertainty 
of $a_e$ may be reduced
by an order of magnitude \cite{vandyck2,gabrielse}.
Further improvement in theory will
enable us to determine $\alpha$ with an uncertainty
of less than 1 part in $10^9$.

Comparison of $\alpha$'s cited above shows that they are
in agreement with each other at the level of $10^{-7}$.
However,  these comparisons must be regarded as
testing theories underlying these measurements,
rather than testing QED. 
Since all these determinations of $\alpha$ are ultimately
based on quantum mechanics, they may be regarded as
testing of the internal consistency of quantum mechanics itself.
It will be of great interest to see whether
the good agreement still holds at the level of
$10^{-8}$ or beyond.

\section{Concluding remark}

Although the first paper Alberto and I wrote together \cite{ks1}
had an embarrassing error, it turned out to be a very productive error.
If this paper did not have the mistake discussed in Sec. 2,
we would not have noticed
the striking cancellation of mass singularities in integrated
quantities, which we emphasized in our subsequent work \cite{ks2}.
I must also point out that we were very lucky
to stumble upon this phenomenon.
This was because 
we were studying the decay process
rather than the  scattering process.
Decay processes have several mass scales
(for instance $m_e$ and $m_\mu$) and it is thus easy to examine
the limit $m_e \rightarrow 0$ while keeping $m_\mu$ finite.
The same cancellation mechanism is also present in scattering
processes.
However, it was not noticed previously because, in a system
with just one mass scale, the mass singularity is not
clearly separated from singularities associated with
threshold behavior or high energy limits.

Although Alberto and I collaborated only for a couple of years
and pursued separate routes afterwards, you will see that much of what
we have done since then have roots in our early collaboration.

This work is supported in part by the National Science Foundation.

\appendix
\section{Communications with Sirlin}

\subsection{Sirlin $\rightarrow$ Kinoshita,~ November 10, 2000}


Dear Tom,

I would like to thank you very much for coming to the Symposium and for
your interesting talk. It was also very nice to meet with your wife and
you after some time! (Although I saw you at the Yang Symposium last year).

Concerning your talk, Massimo Porrati gave me your transparencies and I
noted some discrepancies in the citations to our papers. The problem is
that we wrote a number of papers and letters and it is easy to confuse one
with the other. Here are my observations:

i), ii)  [Errors in the transparencies corrected here.]

%

iii) At the time we corrected our results, I did another check on the
cancellation of mass singularities. I took the corrections for muon decay
for scalar, pseudoscalar and tensor interactions (which we had from the
earlier paper \cite{behrends}), and checked that, 
once the real photon contribution
was corrected, the mass singularity cancels in the integrated spectrum, as
well as the integrated asymmetry. So, from the fact that the cancellation
occurs for the five interactions in muon decay and for the V-A interaction
in beta decay, we had at the time a very strong indication that this was
associated with a powerful theorem, although the proof in the general case
had to wait to your subsequent analysis in \cite{kinoshita2}.

iv) Although it is true that we failed to appreciate the significance of
Feynman's footnote 13, I think that on the particular issue of summing
over polarizations and matching the infrared divergences Feynman was
peculiarly unclear. For example, there is a short book by him, called
$``$Quantum Electrodynamics" (Benjamin/Cummings, 1961), which essentially
contains the material of his 1953 Caltech lectures. On pages 150-151, he
discusses the cancellation of infrared divergences between virtual and
real soft photons in the case of electron scattering by an external
potential, and he only includes the $\ln (K_m /{\lambda_{min}})$
 term from the inner
bremsstrahlung. It seems clear that he only considered the two transverse
directions of polarization in this case and made the same type of
incorrect approximations in the bremsstrahlung integrals as we did in 
\cite{ks1}.

Incidentally, in 1952 Daniel Amati and I were students in
a memorable course in quantum mechanics that Feynman gave in Brazil. At the end of the
course, we asked his guidance about QED and he sent several copies of his
Caltech Lectures. While I waited in Argentina to go to UCLA, I read those
notes. It is rather strange that he did not correct his notes or discuss
this issue in greater detail.

v) While I was a post-doc at Columbia, I received a letter from Feynman,
dated March 7, 1958, which I have kept. The letter was about two issues.

a) It turns out that we had briefly met at some Conference and found out
that we were worried about the same problem, namely the fact that
experimentalists had failed to find the decay 
$\pi \rightarrow e + \bar{\nu}$, and that the
upper-bound in the branching ratio was $< 1/ 10^5$, i.e a factor 12 lower
than the theoretical prediction of the V-A theory! I told him that I was
considering a modification of the theory with the $e$ and $\nu$  coming out at
different space-time points, which would occur, for instance, if there
were a heavy intermediate particle propagating between the two. In the
modern context, this will be the case if leptoquarks actually exist. But of
course the concept of quark and leptoquark had not been proposed at the
time. Feynman told me that he had a different idea: assuming that the self
energy of the electron is purely electromagnetic, he claimed that the most
natural result for the branching ratio was about $3.5 \times 10^{-5}$. 
This still
disagreed with experiment, but there was a factor 4 decrease in the
predicted branching ratio. 
I thought that my approach was extremely
speculative, but I did not see any way out (assuming, of course, that the
experiments were correct). I wrote to T.D. Lee, who was on leave at
Princeton, and he advised me to write a short paper, which I published in
Phys. Rev. 111, 337 (1958). I sent a copy to Feynman, who replied in the letter of
March 7, 1958. He thanked me for the paper and again mentioned his
approach leading to the $3 \times 10^{-5}$ branching ratio. He also added $``$Maybe
experiment is wrong".  

b) In the second paragraph he wrote that his
student Sam Berman had found an error in the correction to the $\rho$-value
that Behrends, Finkelstein, and I had published. He added that the
correction of this error raises $\rho$ by about 0.01 (this is consistent with
the conclusions in our paper \cite{ks2}, and the new Rosenson's value
became $0.68 \pm 0.05$. He also added that he had not seen Crowe's data. If my
memory is right, after the Berman correction, Crowe's value was $0.68 \pm 0.02$.
In any case, it is clear that at the moment Feynman was focusing,
like we, on the corrections to the spectrum. In some sense, we were
fortunate because the corrections to the lifetime became important soon
afterwards (after the implications of the conserved-vector-current
 paper of Feynman and
Gell-Mann became clear), and roughly by that time our results were
corrected.

vi) After we received the Berman paper and corrected our results, I wrote
back to Feynman (on April 29) that Berman's point concerning the need to
include all degrees of polarization associated with a $``$massive photon"
was correct, but we did not agree with the second error he had mentioned,
since it violated the theorem on cancellation of mass singularities. Of
course, at the time we had only a heuristic argument for such
cancellation, rather than your general argument, but the cancellations
were so striking that, I think, both of us were convinced that there was
an underlying theorem. I did not keep a copy of my letter to Feynman (in
those days there were mimeographes rather than Xerox machines), but I do
have a copy of a letter that Berman sent to me, dated May 6, 1958. In the
letter he said that, after reading my letter of April 29 to Feynman, he
rechecked his results and found complete agreement. He also said that, in
preparing the preprint, a copying error was made, resulting in the
spurious term with the mass singularity. Then he thanked me for informing
him of this error, $``$which otherwise might have gone unnoticed". Some time
later, Berman passed through New York and asked me: what is this theorem
you are talking about? If I remember correctly, I told him that at the
time we did not have a general proof, but surely it was a theorem! In
1961, Berman and I overlapped at CERN, became good friends,
 and wrote a nice paper on a number
of subtle points concerning the 
radiative corrections to muon and beta decays.

vii) Sometimes I wonder what would have happened if we had not made the
error. Would we have noticed the cancellation of mass singularities
in integrated quantities, and consequently convinced ourselves of the
existence of an underlying theorem, or would we have missed this most
interesting point?  Because the cancellations are so striking, I think
that we would have found it, but I am not certain. Feynman, with all his
genius, missed it, perhaps because of the copying error! In any case,
I often mention this famous error to my students, and tell them: $``$If
you are going to make a mistake, make a good one and discover a theorem!"

Please, give my best regards to your wife and to Professor Salpeter.

All the best,  ~~~  

Alberto

\subsection{Kinoshita $\rightarrow$ Sirlin,~ November 13, 2000}


Dear Alberto,

Thanks for your informative e-mail.
As you might imagine, I put together the material of my talk
in a hurry and failed to detect errors in citing our papers.
If there is going to be a proceeding of the symposium
(which I strongly hope is the case), these errors will certainly
be corrected.
As a matter of fact, your e-mail contains informations
which will be of interest to readers and historians.
If you are not going to write it by yourself,
do you think it a good idea to attach it to my article
as an Appendix ?
By the way, I remember vaguely that you referred to my
mass singularity paper in your paper on the renormalization
group equation, but I have not found the reference.
Could you give me the proper reference ?
I could then include it in the proceedings.

I am impressed that you keep some letters in your file.
I am rather bad in keeping letters.
I have one letter from Feynman written on a scratch paper,
suggesting that I should write the paper by myself.
It is somewhere in my file, but I have not yet 
located it. I also have Feynman's original two sheet calculation
in my cabinet, but do not know exactly where it actually is.

Please send my best regard to your wife.

Tom

\subsection{Sirlin $\rightarrow$ Kinoshita,~ November 20, 2000}

Dear Tom, 

Thank you very much for your message. Sorry for my delay in
answering: I have been sort of $``$swamped" by urgent departmental matters
and classes. My paper on the renormalization group equation, 
which is rather pedagogical and is based
on considerations of mass singularities, is $``$Mass Divergences and
Callan-Symanzik Equations in Quantum Electrodynamics", Phys. Rev. D5, 2132
(1972).

[Several sentences are omitted here.]

The idea of appending some of the information in my e-mail as an appendix
to your article is fine with me and may be of some historical interest. As
I mentioned, in case anybody is interested, I kept the Feynman and Berman
letters I referred to. What I did not keep, and this is a pity, is a copy
of my reply to Feynman, in which I stated that Berman's additional term
had to be wrong since it violated the cancellation of mass singularities
in the corrections to the lifetime, although, of course, the understanding
of this, in the general case, had to wait for your later work!

All the best,  ~~~  

Alberto


\begin{thebibliography}{99}
\bibitem{haensch}
This value will be improved very soon. 
Private communication from T. H\"{a}nsch.
\end{thebibliography}
\end{document}